\pdfoutput=1
\documentclass[aps,prl,10pt,twocolumn,
noeprint,
nobibnotes, 
footinbib, 
]{revtex4-1}
\usepackage{bm}
\usepackage{amsmath}
\usepackage{amsfonts}
\usepackage{amssymb}
\usepackage{graphicx}
\usepackage[normalem]{ulem}
\usepackage{MnSymbol}
\usepackage{color}
\usepackage[hypertexnames=false]{hyperref}
\usepackage{natbib}

\hypersetup{
    pdfauthor={Pablo M. Perez-Piskunow},
    pdftitle={Hinge Spin Polarization in Magnetic Topological Insulators Revealed by Resistance Switch},
    pdfsubject={Hinge Spin Polarization in Magnetic Topological Insulators},
    pdfkeywords={hinge, spin, magnetic topological insulators}
    }

\newcommand{\startsection}[1]{\emph{#1} \textemdash }
\newcommand{\enquote}[1]{{"{#1}"}}

\begin{document}
\author{Pablo M. Perez-Piskunow}
\email[Electronic address: ]{pablo.perez.piskunow@gmail.com}
\affiliation{Catalan Institute of Nanoscience and Nanotechnology (ICN2), CSIC
and BIST, Campus UAB, Bellaterra, 08193 Barcelona, Spain}
\author{Stephan Roche}
\affiliation{Catalan Institute of Nanoscience and Nanotechnology (ICN2), CSIC
and BIST, Campus UAB, Bellaterra, 08193 Barcelona, Spain}
\affiliation{ICREA--Instituci\'o Catalana de Recerca i Estudis Avan\c{c}ats,
08010 Barcelona, Spain}

\title{Hinge Spin Polarization in Magnetic Topological Insulators
\texorpdfstring{\\}{} Revealed by Resistance Switch}

\begin{abstract}
We report on the possibility to detect hinge spin polarization in magnetic
topological insulators by resistance measurements.
By implementing a three-dimensional model of magnetic topological insulators
into a multi-terminal device with ferromagnetic contacts near the top surface,
local spin features of the chiral edge modes are unveiled.
We find
local spin polarization at the hinges that inverts sign between top and bottom
surfaces. At the opposite edge, the topological state with inverted spin
polarization propagates in the reverse direction.
Large resistance switch between forward and backward propagating
states is obtained, driven by
the matching between the spin polarized hinges and the ferromagnetic contacts.
This feature is general to the ferromagnetic, antiferromagnetic and
canted-antiferromagnetic phases, and enables the design of spin-sensitive
devices, with the possibility of reversing the hinge spin polarization of the
currents.
\end{abstract}
\maketitle

\startsection{Introduction.}
The recent discovery of intrinsic magnetic topological insulator (TI)
multilayered ${\rm MnBi}_{2}{\rm Te}_{4}$ ~\cite{Otrokov2019,Klimovskikh2020}
has boosted the expectations for more resilient quantum anomalous Hall
effect~\cite{Haldane1988,Chang2013,Deng2020,Hirahara2020,Estyunin2020,Wimmer2020,Bonbien2021}
and observability of axion insulator states~\cite{Zhang2019,Liu2020,Giustino2021}.
The material platforms to realize the quantum anomalous Hall (QAH) phase can be
classified, in a broad sense, in two- and three-dimensional systems.
The former includes
monolayer materials and high-symmetry models with spin-orbit coupling and magnetic
exchange~\cite{Qiao2010,Tse2011,[{An alternative route to realize the QAHE is to include strong interactions, leading to orbital magnetization, see }]FanZhang2011,*Chen2019a,*Serlin2020}.
The latter is the case of three-dimensional magnetic TIs, usually realized in thin-films and few-layers systems, including
magnetically doped TIs~\cite{Xu2015,Qi2016},
proximitized TI surfaces with a magnetic
insulator~\cite{Hou2019,Mogi2019}, and the Chern insulator phase of
$\text{MnBi}_2\text{Te}_4$~\cite{Otrokov2019,Liu2020}.
The distinction that arises in three-dimensional magnetic TIs is that
the topological nature comes from contributions
from two Dirac-like surfaces that, upon the introduction of
magnetization field throughout the material, become massive with opposite
effective masses~\cite{Yu2010,Chu2011}.
Despite the three-dimensional nature of magnetic TIs, they are often analyzed
near the surface, as effective two-dimensional systems.

However, compared to their two-dimensional counterparts, three-dimensional
magnetic TIs present a higher level of complexity that
reflects in layer-to-layer magnetic exchange and termination-dependent surface
states, which ultimately dictate
the nature and properties of surface magnetism and of topological edge
states~\cite{Tokura2019,Wu2020,Nevola2020}.
The spin texture of topological edge states in both the quantum spin Hall and
quantum anomalous Hall (QAH) regimes is usually perpendicular
to the material's surface, limiting the possibility for magnetic-sensitive
detection or further spin manipulation protocols~\cite{[{For tunable spin polarization in the QAHE see }]RXZhang2016}.
The effective two-dimensional models of these materials are often highly
symmetric and may overlook the sublattice and spin degree of freedom. However, spin textures~\cite{[{A 2D model including sublattice and spin can yield in-plane spin polarization, see }]Wu2014}, spin Hall conductivity~\cite{Costa2020},
and local spin polarization~\cite{Plekhanov2021} provide great
insight into the special topological phases that can arise in topological
superconductors and boundary-obstructed
TIs~\cite{[{A study of the quantum spin Hall effect in magnetic topological insulators indicates the presence of hinged quantum spin Hall states, characterized by a non-trivial spin-Chern number }]Ding2019,*[{
However similar, our model cannot be characterized by the spin-Chern number due to the
large SOC and non-conservation of the spin, see }]Prodan2009,*[{ and }][{ Instead, our model
is characterized by a non-zero Chern number.}]Monaco2020,Khalaf2019}.
By reducing the symmetry constrains, new spin textures can develop, such as hidden
spin polarization~\cite{Zhang2014} and canted spin
textures~\cite{Shi2019,Vila2020,Garcia2020}.
In presence of a uniform~\cite{Zhang2013a} or alternating~\cite{Plekhanov2020}
Zeeman field, several models of magnetic layers exhibit
high-order topological phases and cleavage-dependent hinge
modes~\cite{Zhang2013a,Varnava2018,Trifunovic2020,Tanaka2020,Zhang2020,Plekhanov2020,[{For a similar model describing magnons see }]Mook2020}.
Thus, a detailed study of the spin features on a spinful three-dimensional
model of the QAHE realized in magnetic TIs multilayers is missing.

In this Letter, we use the generic Fu-Kane-Mele (FKM) model for three-dimensional
topological insulators~\cite{Fu2007a} and introduce exchange terms to describe
both ferromagnetic (FM) and antiferromagnetic (AFM) multilayered TIs.
Contrary to ordinary spin-$z$ polarization of edge states in the QAH regime, the
model exhibits an in-plane \emph{hinge spin polarization} (HSP) which becomes
apparent (and observable) in a specific device setup.
Indeed, the topological states are characterized by an in-plane HSP
perpendicular to both the current flow and the sample
magnetization direction. The in-plane polarization reverses sign
along the vertical direction, between the top and bottom surfaces.
By using efficient quantum transport simulation methods~\cite{Groth2014}
implemented into a three-dimensional multi-terminal device, such peculiar
local spin polarization is shown to give rise to a giant resistance
switching (or \emph{spin valve}) triggered upon either inverting the
magnetization of the sample, varying the polarization of the magnetic detectors,
or reversing the current direction.
The appearance of HSP in the QAH regime is rooted in the
chiral-like~\cite{Wu2014,Zhang2020}
symmetries of the lattice, and on the half-quantization of the topological
charge at the surfaces~\cite{XLQi2008,Chu2011,Gu2020,RLu2020,[We verified the half-quantized topological charge at the surfaces using the method described in ]Varjas2020}. Therefore, the HSP fingerprints are
highly robust to Anderson-type of energetic disorder, and to structural edge
disorder.

\startsection{Hamiltonian of the three-dimensional magnetic TI.}
The magnetic TI is described by a three-dimensional
(diamond cubic lattice) FKM
Hamiltonian~\cite{Fu2007,Fu2007a,Soriano2012},
with magnetic layers modelled by an exchange coupling term that
well captures the effect of magnetic impurities~\cite{Liu2009}  or magnetic
layers ~\cite{Zhang2013a,Wang2013}.
To simulate a multilayer FM or AFM magnetic TI we tune the orientation of the
magnetic moments per layer.
The FKM lattice vectors are
$\bm{a}_1=\left(1/2,\,-\sqrt{1/3}/2,\,\sqrt{2/3}\right)$,
$\bm{a}_2=\left(0,\,\sqrt{1/3},\,\sqrt{2/3}\right)$, and
$\bm{a}_3=\left(-1/2,\,-\sqrt{1/3}/2,\,\sqrt{2/3}\right)$; each unitcell has
two sublattices: $A$ with $\bm{0}$ offset, and $B$ with offset
$\bm{d}_4=(0,\,0,\,\sqrt{3/2}/2)$. The other first neighbors of $A$ sites are
at relative positions $\bm{d}_q=\bm{d}_4-\bm{a}_q$ for $q=1,2,3$.
The full Hamiltonian reads
\begin{equation}\label{eq:ham}
\begin{split}
 \mathcal{H}_0 &= \sum_{\langle i,j \rangle} \sum_{\alpha}
 c^\dagger_{i,\alpha} t_{ij} c_{j,\alpha}
 ;\quad
  \mathcal{H}_Z =
  \sum_{i,\alpha, \beta}c^\dagger_{i, \alpha}
 \left[\bm{m}_i\cdot\bm{s}\right]_{\alpha, \beta}
 c_{i, \beta}\\
 \mathcal{H}_{SO} &=i\frac{8\lambda_{SO}}{a^2}
 \sum_{\llangle i, j \rrangle} \sum_{\alpha, \beta}
 c_{i,\alpha}^\dagger \left[\bm{s}
\cdot(\bm{d}_{ij}^1\times\bm{d}_{ij}^2)\right]_{\alpha, \beta}
 c_{j, \beta} \\
 \mathcal{H} &= \mathcal{H}_0 + \mathcal{H}_{SO}  + \mathcal{H}_Z,
\end{split}
\end{equation}
with latin indices for lattice sites, and Greek indices for spin value in
the $s_z$ basis.
The Zeeman magnetization vector $\bm{m}_i$ may depend on the layer of the
orbital $i$, and $\bm{s}$ is a vector of Pauli
matrices acting on the spin degree of freedom.
The parameter $\lambda_{SO}$ denotes the spin-orbit coupling strength, while
$t_{ij}$ describes the nearest neighbors coupling between sites $i$ and
$j$, and takes different values $t_q$ with $q=1...4$ depending on the direction
$\bm{r}_j - \bm{r}_i = \bm{d}_q$. As
described in~\citet{Fu2007a}, the isotropic case $t_q=t$ defines a
multicritical point. Adding anisotropy $t_q=t$ for $q=1, 2, 3$
and $t_4 > t$, sets the phase to a strong TI characterized by a
non-trivial $\mathcal{Z}_2$ invariant.
We tune the parameters to the strong TI
phase with $t_4=1.4\,t$ and $\lambda_{SO}=0.1\,t$~\cite{Note1}.
%
The FKM model can be interpreted as a stack of coupled Rashba layers, with
alternating Rashba field~\cite{Pershoguba2012, Plekhanov2020}. In absence of
Zeeman field the strong TI phase is the three-dimensional realization of the
Shockley model~\cite{Pershoguba2012}, hosting sublattice
polarized surface states.
The magnetic moments per layer describe the AFM
(alternating magnetization between layers $\bm{m}_i=\pm\bm{m}$)
or FM (constant magnetization $\bm{m}_i=\bm{m}$) coupling between layers.
In a slab geometry perpendicular to the $z$ axis, a Zeeman exchange coupling
field $\bm{m}=0.05\,t\,\hat{z}$ opens a gap on the
surface states, and sets the QAH phase described by a non-trivial Chern
number~\cite{Liu2016}.

\begin{figure}
  \includegraphics[width=\linewidth]{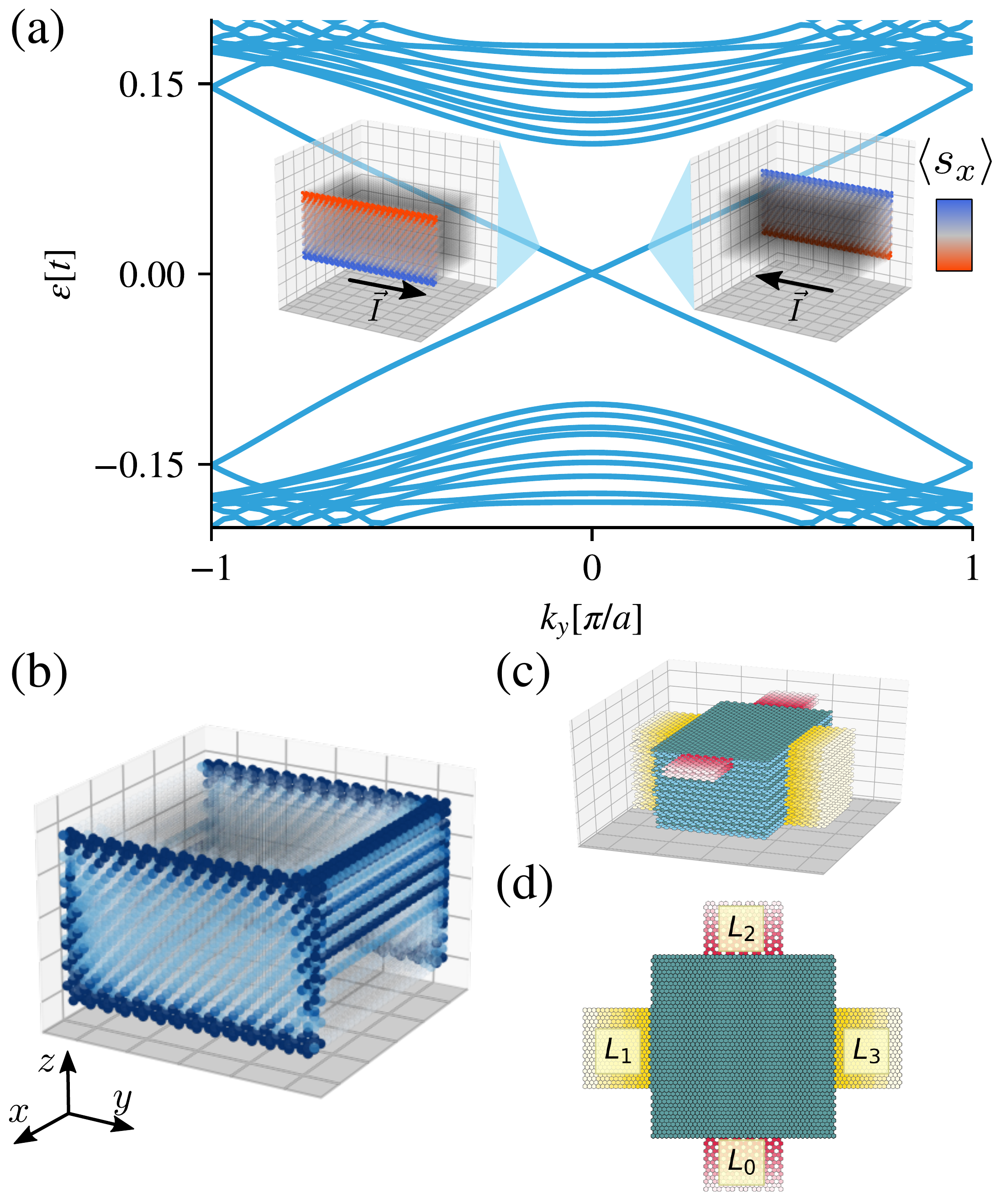}
  \caption{\label{fig1}
    Magnetic TI in the FM phase, $\bm{m}=0.05t\,\hat{z}$.
    a) Dispersion relation of a slab geometry infinite in the $y$ direction.
    The left (right) inset depicts the local spin density of states $\langle
s_x\rangle$ of the edge state at $k_y=-0.1\pi/a$ ($k_y=0.1\pi/a$). The edge
state covers the sidewall of the slab and propagates to the right (left).
    b) Local density of states of a finite square slab. The edge state
    circulates around the sample, covers the side surfaces perpendicular to
    $\hat{x}$, and propagates along the top or bottom hinges of the side
    surfaces perpendicular to $\hat{y}$.
    c) Side view of transport setup geometry: metallic leads connect to the
whole walls at both ends of the slab (golden color), and ferromagnetic leads
connect to the sidewalls only near top hinge (red color).
    d) Top view and reference numbering of the leads on the transport setup.
  }
\end{figure}

We present the main electronic and spin characteristics of the
magnetic topological insulator model in Figure~\ref{fig1}.
The details of the edge modes vary with the geometric design. For
a heterostructure infinite along the $y$-direction but finite in both
other directions, we obtain the usual linear energy dispersion of topological
edge states seen in Fig.\ref{fig1}-a). These states cover the whole side surface
of the stack (wall-states) with a very large electronic density at the hinges.
Interestingly, the projected local spin density of the wall-states is seen to be
dominated by the $\langle s_{x}\rangle$ value near the hinges. The \emph{hinge
spin polarization} (HSP) switches sign between opposite surfaces.
Furthermore, the HSP changes sign for the back-propagating states, located
at the opposite walls (see insets).
On a finite slab, Fig.~\ref{fig1}-b), the nature of the chiral states becomes
richer, with the emergence of hinge states for certain surface cleavage
orientations, a property predicted for \emph{M\"obius}
fermions~\cite{Zhang2013a,Zhang2020,Plekhanov2020}.
The \emph{M\"obius} fermions phase depends on the ferromagnetic interlayer
exchange, and appears in the FM and canted-AFM phase on crystalline canting
directions.
Conversely, the HSP is robust and appears in all phases, that is: FM, AFM,
and canted-AFM, irrespective of the canting angle, as long as there is a
$z$-component of the net magnetization.
We next explore the
possible fingerprints of such anomalous spin features on quantum transport in
the QAH regime.

\startsection{Multi-terminal spin transport simulations.}
To analyse the spin transport in the QAH regime, we use the Kwant
software package~\cite{Groth2014} to build the three-dimensional
model, and implement a multi-terminal device configuration, shown in
Figs.~\ref{fig1}~c), and d).
We perform charge transport simulations of a central scattering region connected
with metallic and ferromagnetic leads.
The interplay between the states available for transport in the leads and in
the scattering region has a central role.
The leads $L_1$ and $L_3$ are the metallic leads (golden color).
They are fully contacting the left and right sides of the slab (all spin
projections).
The ferromagnetic leads $L_0$ and $L_2$ (red color) located on the sides
only contact the upper part of the device near the top hinge~\cite{[{A similar device can be envisioned to measure the Axion insulator phase }]RChen2020}.
They carry electrons with only one spin polarization: $(s_x,\downarrow)$.
In this way, these contacts couple with the edge state in the region of maximal
local spin polarization.

\begin{figure}
 \includegraphics[width=\linewidth]{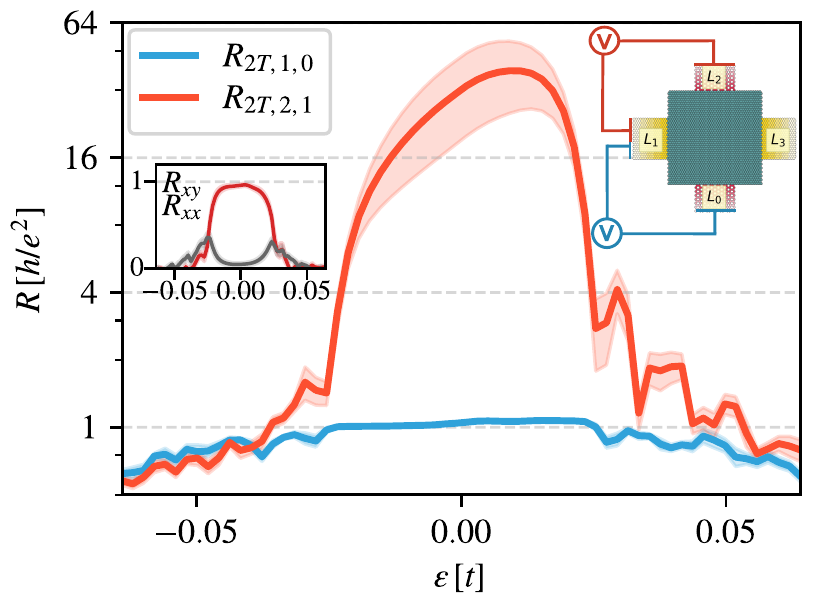}
 \caption{\label{fig2}
 Transport simulations of a FM slab ($\bm{m}=0.02t\,\hat{z}$)
 between metallic leads and ferromagnetic leads
 with spin down $(s_x,\downarrow)$ polarization. The shaded regions depict the
 standard error of considering $10$ Anderson disorder realizations of strength
 $d=0.04\,t$.
 A 4-terminals device allows to measure the distinct resistance profiles.
 The two-terminal resistance setup between leads $L_0$ and $L_1$ is depicted on
 the left inset and the blue curve is the resistance profile.
 The right inset shows the resistance setup between leads $L_1$ and $L_2$,
 with resistance profile in red. In the former case the ferromagnetic lead
 polarization matches the top HSP, and in the latter case
 and in the second case the local spin polarization is opposite.
 }
\end{figure}

The expected resistance measurements for
the QAHE are shown on the inset of Fig.~\ref{fig2}.
We use the notation $R_{ij, kl}$, for the resistance
measured from passing current between terminals $i$ and $j$, and measuring the
voltage drop between terminals $k$ and $l$. The two-terminals (2T) resistance
$R_{kl,kl}$ is noted $R_{2T,kl}$.
The typical values of Hall
resistance $R_{xy}=R_{13,20}$ and the longitudinal resistance $R_{xx}=R_{01,23}$
of a QAH
insulator~\cite{Buttiker1988a,Buettiker1986} take the quantized values
$R_{xy} = \frac{1}{C} h/e^2$, where $C$ is the Chern number,
and vanishing $R_{xx}$ inside the gap.
The two-terminal resistance $R_{2T}=h/e^2$ is also quantized for
perfect tunneling between the leads and the scattering region~\cite{Note2}.
Such is the case of the matching ferromagnetic lead.
The matching or mismatching between the spin current
carried by the leads and the spin polarization of the edge states gives rise
to a remarkable resistance switch, as seen in Fig.~\ref{fig2}. The 2T
resistance in the matching case is quantized inside the
topological gap, while in the mismatching case the resistance
increases by more than one order of magnitude.

To test the robustness of the 2T resistance switching
effect, we introduce different types of disorder. First, we consider
the impact of structural disorder---vacancies near the sidewalls of the slab.
This disorder is detrimental to the formation of well-defined
HSP, which only occur for wall-states at crystalline edges.
Nevertheless, is relevant for predictions on experiments, since the side walls
of material samples have edge disorder.
We find that the HSP effect survives to structural disorder, up to $\%5$
vacancies~\cite{Note1}.
Next, we simulate Anderson disorder by adding an onsite energy $d\chi$,
where $\chi$ is a random variable with normal distribution on $[-0.5, 0.5)$.
We find robustness of the HSP up to $d$ much larger than the magnetization
strength. In Fig.~\ref{fig2}, we use $d=2\lvert\bm{m}\rvert=0.04\,t$, and
average the resistance curve over $10$ disorder realizations. We see that spin
transport measurements can still distinguish the peculiar spin texture of the
edge states.

The fact that Anderson disorder and structural disorder show the
resistance switch is crucial in establishing the robustness of our results.
The limit mismatching case, where the edge state a ferromagnetic lead
are completely decoupled from the transport setup, results in voltage probes
that have zero transmission probability to any other leads, leaving a floating
probe with an arbitrary value of the chemical potential and the
voltage~\cite{Note3}.
%
However, in our case the ferromagnetic leads are not fully disconnected when the
spins do not match, rather they are weakly connected.
Even though the value of the 2T resistance is sensitive to the
details of the weak coupling, seen on the large standard error in
Fig.~\ref{fig2}, the trend is clear.
In a QAH thin-film contacted on its lateral sides with ferromagnetic leads,
we can selectively get, either full transmission, or blocking of the edge state
transport.
Such phenomenon is sensitive to the direction of the magnetization of
the ferromagnetic leads, the direction of the current, and the net magnetization
of the sample.

\begin{figure}
 \includegraphics[width=\linewidth]{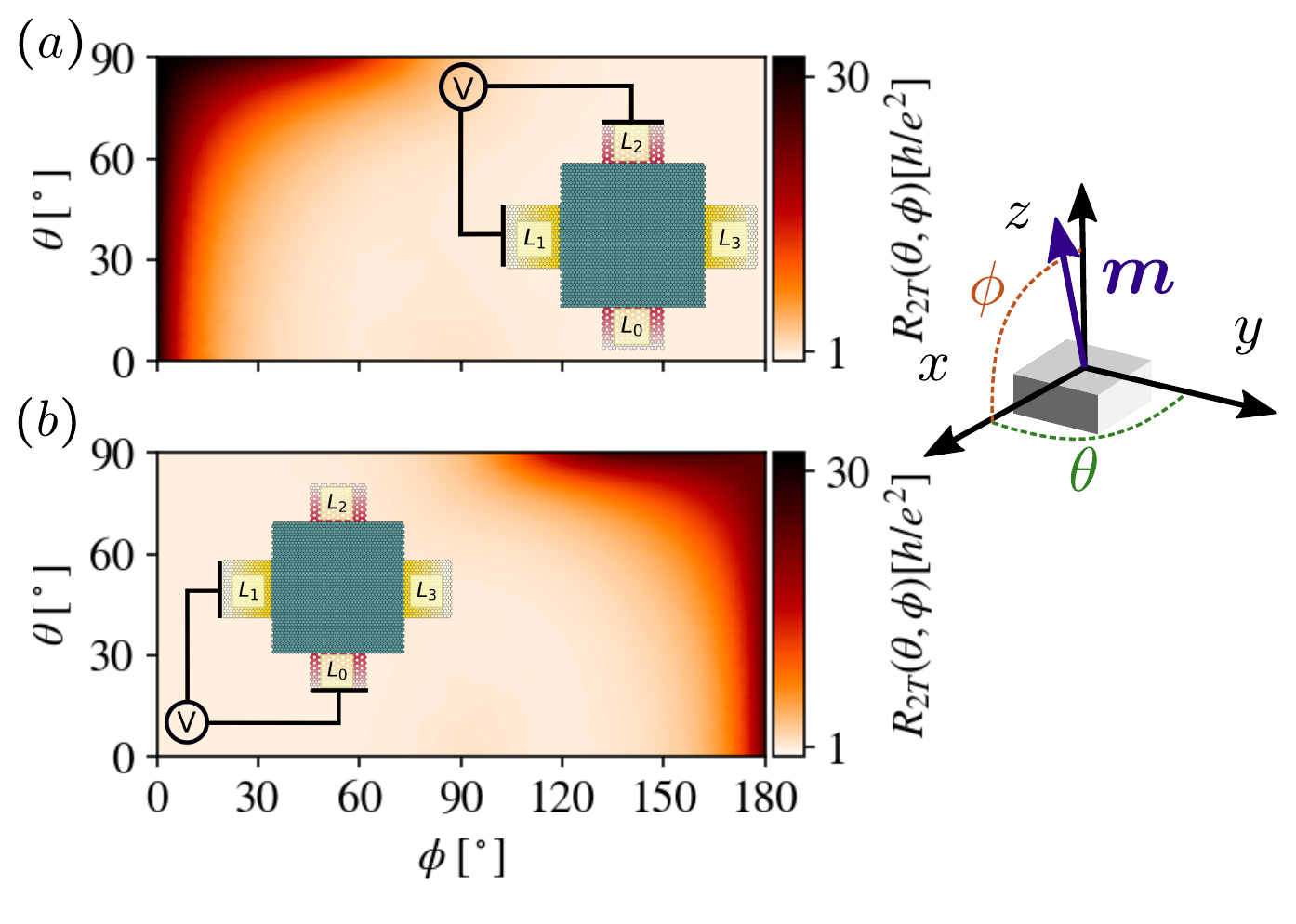}
 \caption{\label{fig3}
 2T resistance at $e_F=0$ of a magnetic TI slab in the FM phase, with
 magnetization $\lvert\bm{m}\rvert=0.02\,t$ (see right inset).
 In a) the two-terminal resistance is measured between leads $L_1$ and $L_2$;
 in b) the resistance is measured between leads $L_0$ and $L_1$.
 The plots in a) and b) are almost identical by reflecting around
 $\phi=90^\circ$, where the magnetization component along the $z$-axis
 changes sign.
 The mismatching configurations $\phi=0^\circ$ in a), and $\phi=180^\circ$
 in b) shows the largest resistance, that slowly decreases when rotating the
 magnetization angle.
 }
\end{figure}

Another experimentally relevant analysis is to explore the resistance switch
for different directions of the magnetization $\bm{m}$ of the slab in the FM
phase. Figure~\ref{fig3} shows two measures of 2T resistance, as in
Fig.~\ref{fig2} at the charge neutrality point, for different directions of the
Zeeman exchange field $\left(\cos{\theta}\cos{\phi}\hat{x} +
\sin{\theta}\cos{\phi}\hat{y} + \sin{\phi}\hat{z}\right)$ (see right inset).
At low $\phi$ angles ($\bm{m}$ pointing mostly towards $+\hat{z}$)
the configuration $R_{2T,1,2}$ in a) shows large resistance,
while $R_{2T,0,1}$ in b) is close to the quantized value $h/e^2$.
When sweeping the magnetization to the inverse direction (towards $-\hat{z}$)
at $\phi~180^\circ$, the roles of a) and b) reverse, giving a clear signature
of the highly spin-polarized hinges and of the spin-dependent matching and
mismatching with the ferromagnetic leads.
In the middle of both extremes where $\phi=90^\circ$, the magnetization lies
in the plane of the slab and does not open a gap on the top and bottom surfaces.
At intermediate angles, we note that the resistance switch is more robust for
$\theta=90^\circ$, where $\bm{m}$ tilts towards $\hat{y}$,
the transport direction and edge direction that the FM leads contact.

\begin{figure}
 \includegraphics[width=\linewidth]{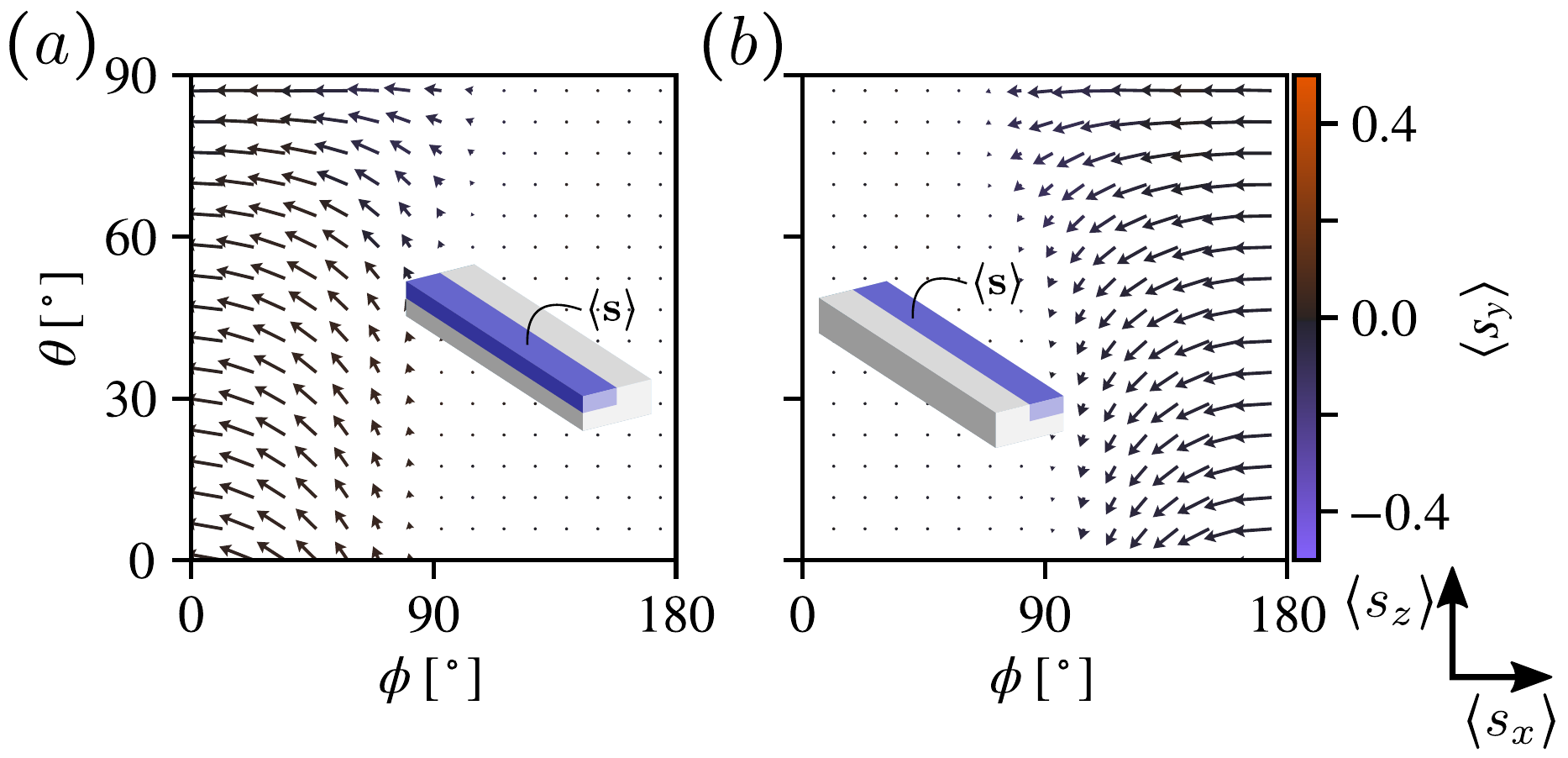}
 \caption{\label{fig4}
 Magnetic TI slab with the same parameters as in Fig.~\ref{fig3}.
 HSP of the edge state at $k_y=-0.02\pi/a$ and positive energy
 (propagating in the $\hat{y}$ direction) for a wide range of magnetization
 angles $(\theta, \phi)$.
 The arrows show the components of $\langle {s}\rangle$ projected on the top-
 left (-right) hinges of an infinite slab along the $y$-axis. The direction of
 the arrows gives the component in the
 $(\langle s_x\rangle, \langle s_z\rangle)$ plane, and the color of the
 arrows is the component in $\langle s_y\rangle$.
 a) Propagating edge states on the top-left hinge are only present when the
 magnetization has a positive projection along $z$, thus
 $\phi \lesssim 90^\circ$.
 b) Propagating edge states on the top-right hinge are only present when the
 magnetization has a negative projection along $z$, thus
 $\phi \gtrsim 90^\circ$.
 }
\end{figure}

The HSP of the edge states is a good proxy to predict the
switch in resistances that is measured in the device shown in
Fig.~\ref{fig3}.
We obtain the spin projection of the forward
propagating edge states on the top half of an infinite slab in the $\hat{y}$
direction, and finite in the $xz$ plane, see insets of Fig.~\ref{fig4}.
At momentum $k=-0.02\pi/a$
we select the positive eigenvalue inside the topological gap,
similar to the states shown in the insets of Fig.~\ref{fig1}-a).
Panels a) and b) of Fig.~\ref{fig4} show finite length arrows that indicate the
spin density and components in the $(\langle s_x\rangle, \langle s_z\rangle)$
plane of the forward propagating state, while the color represents the mostly
null $\langle s_y\rangle$ component.
A vanishing arrow length (a point in the plot) indicates that there is no net
spin density at that region enclosing that hinge~\cite{Note4}.
%
When the system is in the topological phase ($\phi\nsim90^\circ$),
there is electronic density in one edge \emph{or}
the other, and spin density near the hinge (a finite arrow).
Accordingly, we see that panel b) complements perfectly panel a).
In both cases the HSP direction changes with the
magnetization angle, giving a notch to control the matching or mismatching
cases in a transport setup.

\startsection{Conclusions.}
We have demonstrated that the edge states in thin-film ferromagnetic and
antiferromagnetic TIs host HSP,
spin polarized states at the hinges,
which leads to a large resistance switch.
The HSP of the edge states is in-plane, but the sign depends on the propagation
direction and the magnetization of the sample.
For a crystalline edge direction, the local
spin polarization reverses across the vertical direction.
Thus, the HSP inverts across the vertical direction, and switches sign for
the opposite current direction.
Carefully engineering ferromagnetic contact leads in a transport
setup, allows us to obtain a \emph{giant resistance (spin valve effect)} upon
reversing the current direction or, conversely, tuning the total magnetization
of the sample.
The $\langle s_x\rangle$ component of the spin direction in Fig.~\ref{fig4} a)
and b) can be directly translated to the resistance values found in
Fig.~\ref{fig3} a) and b). This highlights that the resistance switching
mechanism, once established, can be used to gain insight about the
magnetization of the sample.

We finally observe that the fact that FM and AFM topological insulators are able
to host maximally spin polarized currents along crystalline hinges
opens new avenues to
implement disruptive proposals using Axion and magnetic TIs to manipulate
dislocation, hinge, and edge states~\cite{Varnava2018}, with the additional
value of spin polarization features.

\begin{acknowledgments}
\startsection{Acknowledgments.}
We thank Sergio O. Valenzuela, David Soriano, and Aron W. Cummings for fruitful
discussions.
We acknowledge the European Union Horizon 2020 research and innovation
programme under Grant Agreement No.  824140 (TOCHA, H2020-FETPROACT-01-2018).
ICN2 is funded by the CERCA Programme/Generalitat de Catalunya, and is
supported by the Severo Ochoa program from Spanish MINECO (Grant No.
SEV-2017-0706).
\end{acknowledgments}

\onecolumngrid
\clearpage
\appendix

\newcommand{\hbAppendixPrefix}{S}
\renewcommand{\thefigure}{\hbAppendixPrefix\arabic{figure}}
\setcounter{figure}{0}
\renewcommand{\thetable}{\hbAppendixPrefix\arabic{table}}
\setcounter{table}{0}
\renewcommand{\theequation}{\hbAppendixPrefix\arabic{equation}}
\setcounter{equation}{0}
\renewcommand{\thepage}{\hbAppendixPrefix\arabic{page}}
\setcounter{page}{1}
\renewcommand{\thesection}{\hbAppendixPrefix\arabic{section}}
\setcounter{section}{1}

\begin{center}
 \textbf{\large Supplemental Material}
\end{center}

\section{Spin texture of the Fu-Kane-Mele model with Zeeman field}

\begin{figure}[b]
    \includegraphics[width=0.98\linewidth]{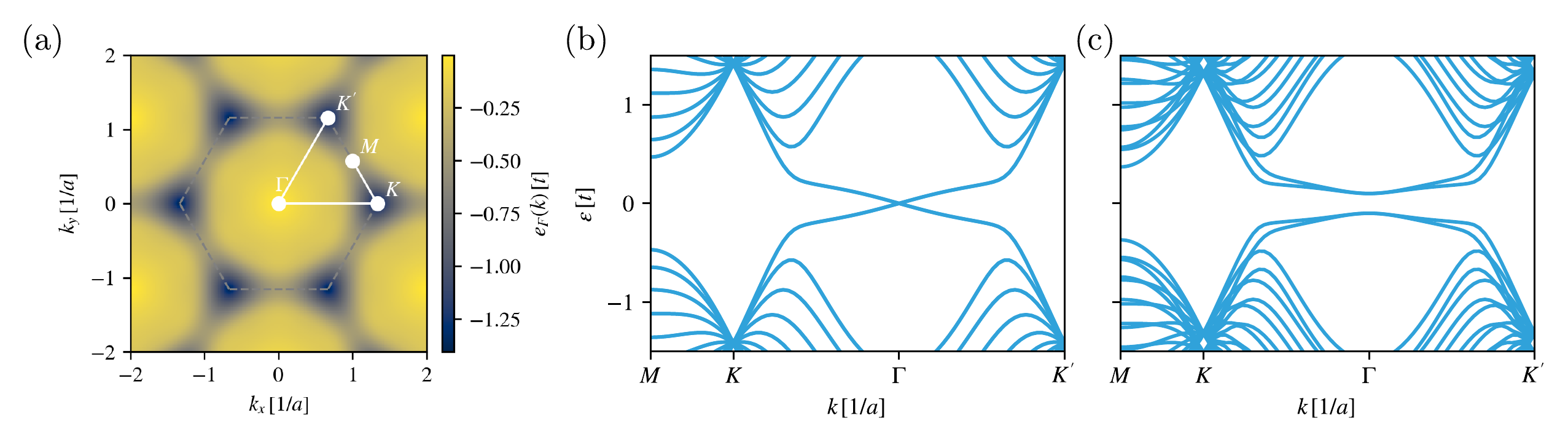}
    \caption{\label{kpath}
    (a) Fermi level resolved in $\bm{k}$ space of a FKM slab with $12$ layers.
    The Dirac point appears at the $\Gamma$ point. The dotted gray line delimits the Brillouin zone, and high-symmetry points are denoted with white dots. The k-path is traced with a white line connecting the high-symmetry k-points.
    The band-structure following the k-path is depicted in (b) for zero Zeeman field, and in (c) for finite Zeeman field ($\lvert\bm{m}\rvert=0.1\,t$).
    }
\end{figure}

The Fu-Kane-Mele~\cite{Fu2007_SM,Fu2007a_SM,Soriano2012_SM,Pershoguba2012_SM} (FKM) model is a versatile model of
topological insulators that
can reproduce $1$ or $3$ Dirac cones on the surface, depending on the surface
termination.
If Fig.~\ref{kpath} we show the case of one Dirac cone at each surface of a
two-dimensional slab, finite in the $z$-direction.
We obtain one Dirac point (per surface) located  at the $\Gamma$ point
by setting the surface termination to $A(B)$ at the top (bottom)
surface, and $t_4=t'$, $t_i=t$ for
$i=1,2,3$, with $t'>t$.
With a small but finite Zeeman field, the Dirac cones at each surface become
massive, giving rise to the QAH phase.

\begin{figure}
    \includegraphics[width=0.8\linewidth]{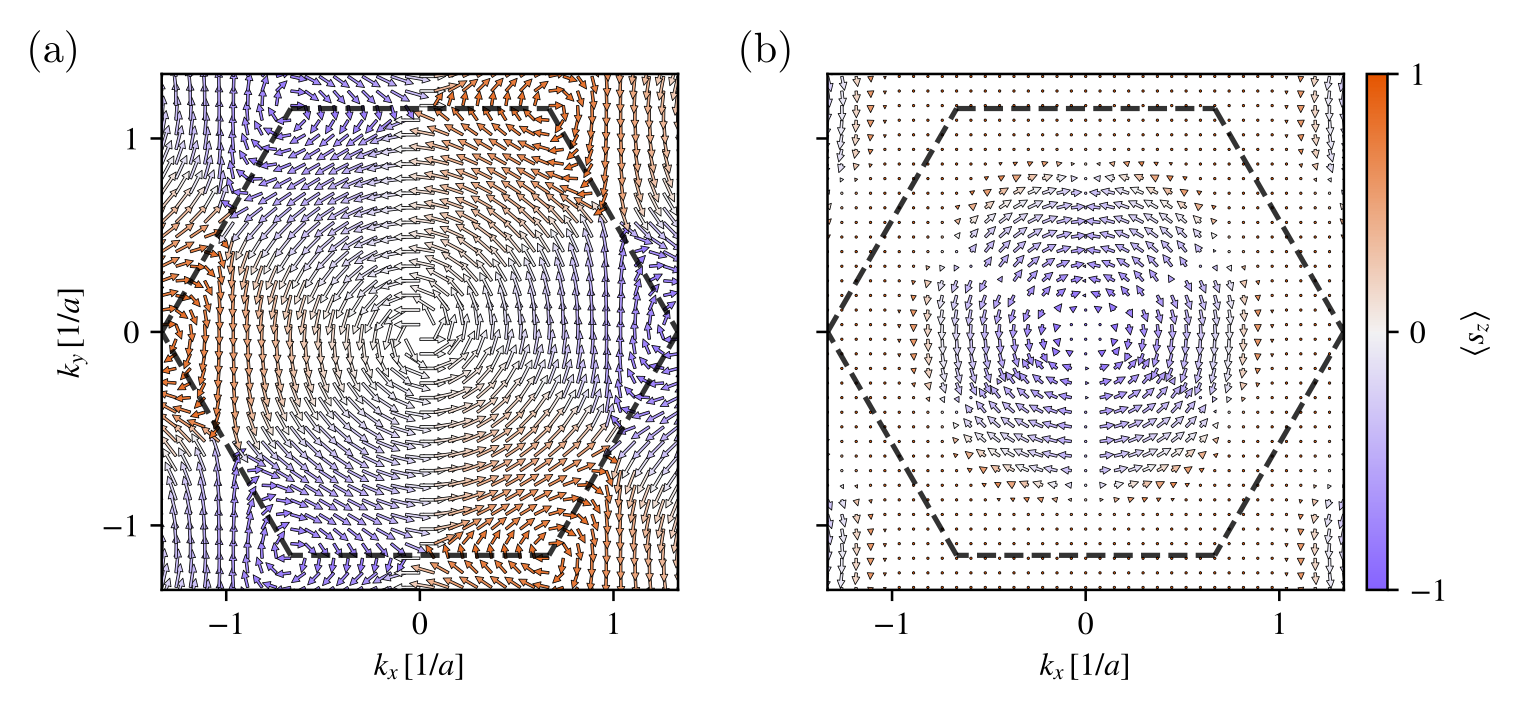}
    \caption{\label{texture}Spin texture of the FKM model. (a) without Zeeman field, and (b) with Zeeman field $\lvert\bm{m}\rvert=0.05\,t$.}
\end{figure}

The spin texture in Fig.~\ref{texture}-(a)
of the FKM model compares qualitatively
well with that of
$\text{Bi}_2\text{Te}_3$~\cite{Liu2010a_SM}.
The model reproduces the time-reversal symmetry $\mathcal{T}$,
and three-fold rotation symmetry $\mathcal{C}_3$, while the bulk gap
can be tuned by $\Delta t=t'-t$ and the spin-orbit coupling strength
$\lambda_{SO}$.

In Fig.~\ref{texture} (b) we show the spin texture in the
Quantum Anomalous Hall phase.
A crystalline termination along the $y$ direction
(and $3$-fold rotation symmetric directions) hosts bulk
states with zero net spin polarization in the $xy$-plane.
The edge states inherit the same pattern,
giving a zero net spin polarization in the $xy$-plane.
This is also a consequence of one edge mode hosting two hinge spin polarizations
near the top and bottom surfaces, that cancel each other out exactly.

\section{Spatial profile of the edge-states}

The edge-states carrying hinge spin polarization (HSP) are \emph{wall-states}
(as opposed to hinge-states). They cover the whole side of the slab, and host
finite electronic density throughout. This means that the top and bottom HSP
form part of the same state, and injecting current on either, will propagate
the electronic density to the other hinge. As a consequence of the
non-conservation of the spin in a system with strong spin-orbit coupling (SOC),
even if injecting perfectly spin-polarized current, when propagating through
the slab, the spin-polarization of current will take the values dictated by
the two HSPs.

\begin{figure}[h]
    \includegraphics[width=0.9\linewidth]{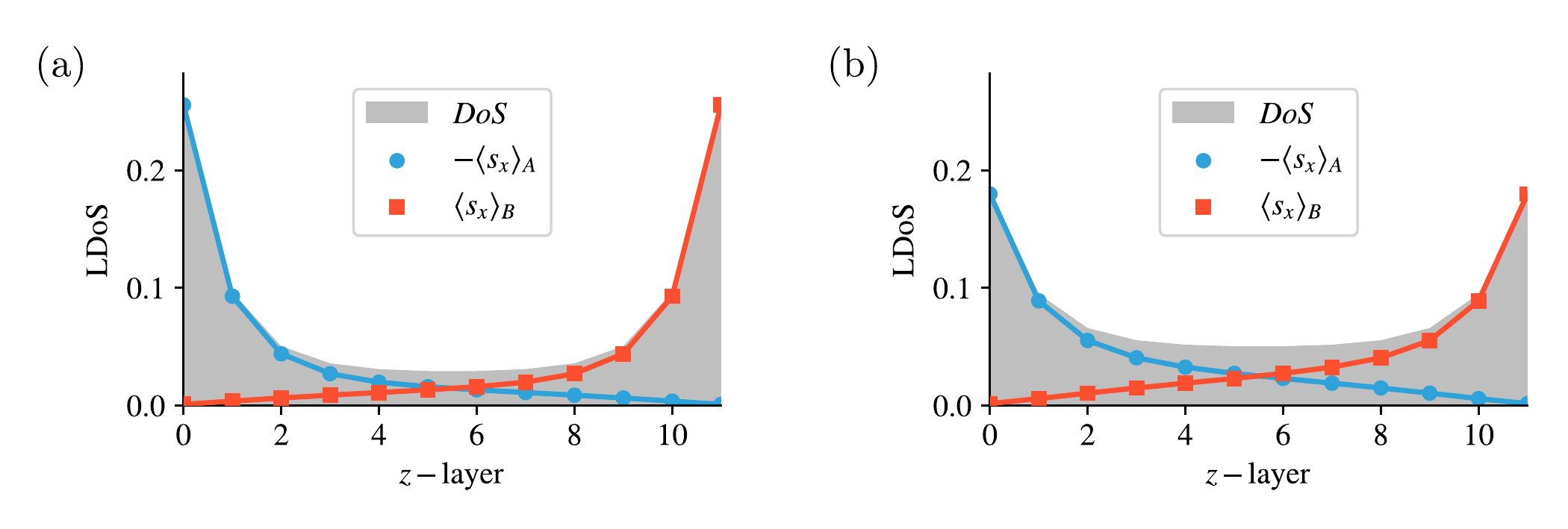}
    \caption{\label{spin_profile} Spatial profile of the local density of states
    (LDoS) and spin projection per layer of \emph{one} edge state located at
    $k=0.05\,a^{-1}$ and positive energy. The single edge state has distinct
    sublattice and spin profiles. The spin projection $\langle s_x\rangle$ of
    sublattice $A$ is shown in blue circles, and the one of sublattice $B$ in
    red squares.
    In (a) the magnetization is $\bm{m}=0.05\,t\hat{z}$,
    while in (b) is $\bm{m}=0.1\,t\hat{z}$.
    Note the HSP is sublattice polarized, and that we reversed the sign of
    $\langle s_x\rangle_A$.}
\end{figure}

To demonstrate this, we show in Fig.~\ref{spin_profile} the spatial profile
of one edge-state
with the two HSPs. The markedly high (spin-polarized) electronic density at the
hinges is a result of the half-quantized topological charge. A signature
of the surfaces hosting Dirac states at zero Zeeman field, and these electronic
modes being pushed to the hinge at small but finite Zeeman field.

\section{Structural disorder simulations}

\begin{figure}
    \includegraphics[width=0.49\linewidth]{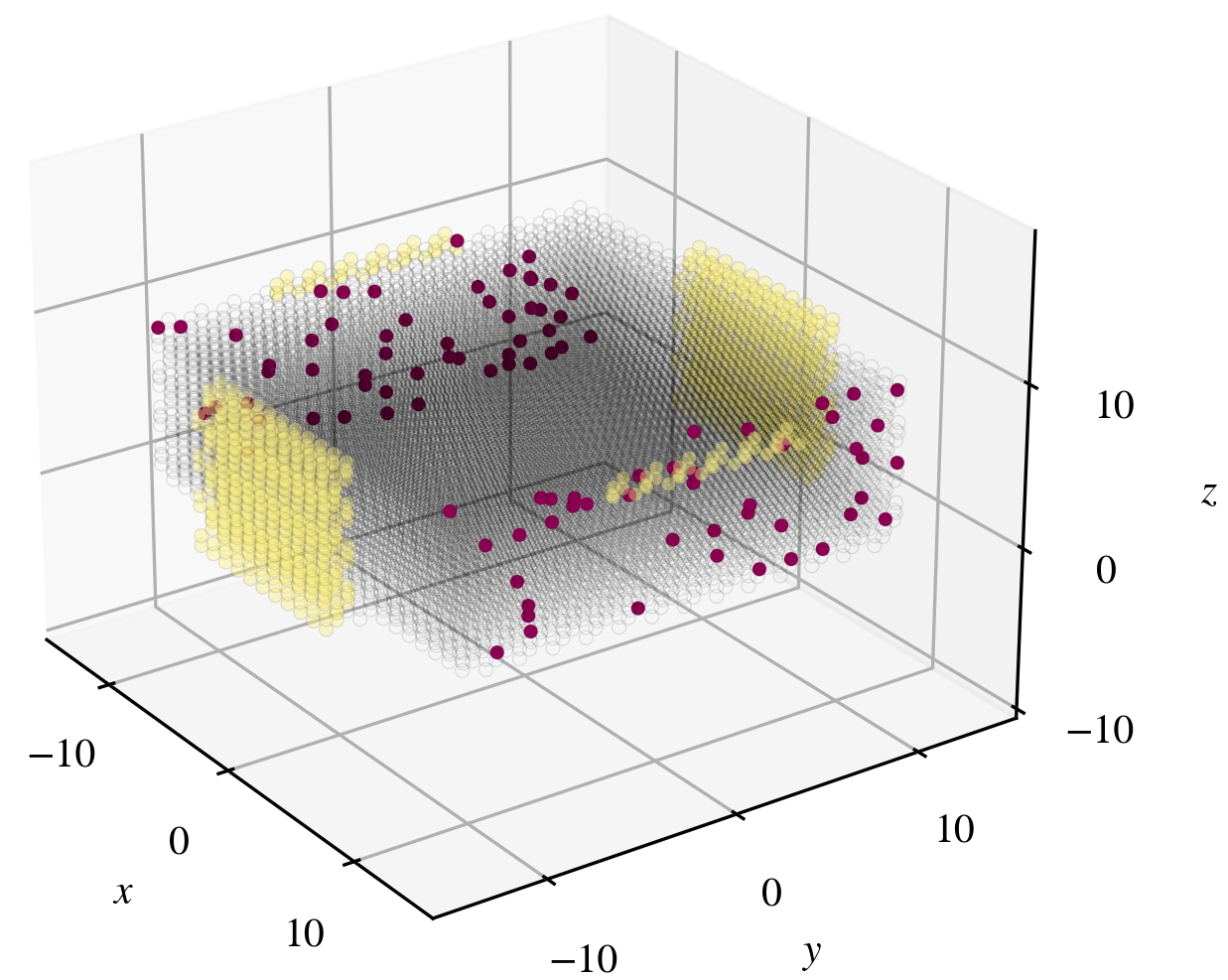}
    \caption{\label{struct_show} One realization of structural disorder at the edges of a slab. The probability of removal is $p=0.05$ for all sites
    within $3$ lattice constants from the edge parallel to the $y$ axis.
    The sites in purple are eliminated from the transport setup. One unitcell of
    the leads connected to the slab are depicted in golden color.}
\end{figure}

\begin{figure}
    \includegraphics[width=0.95\linewidth]{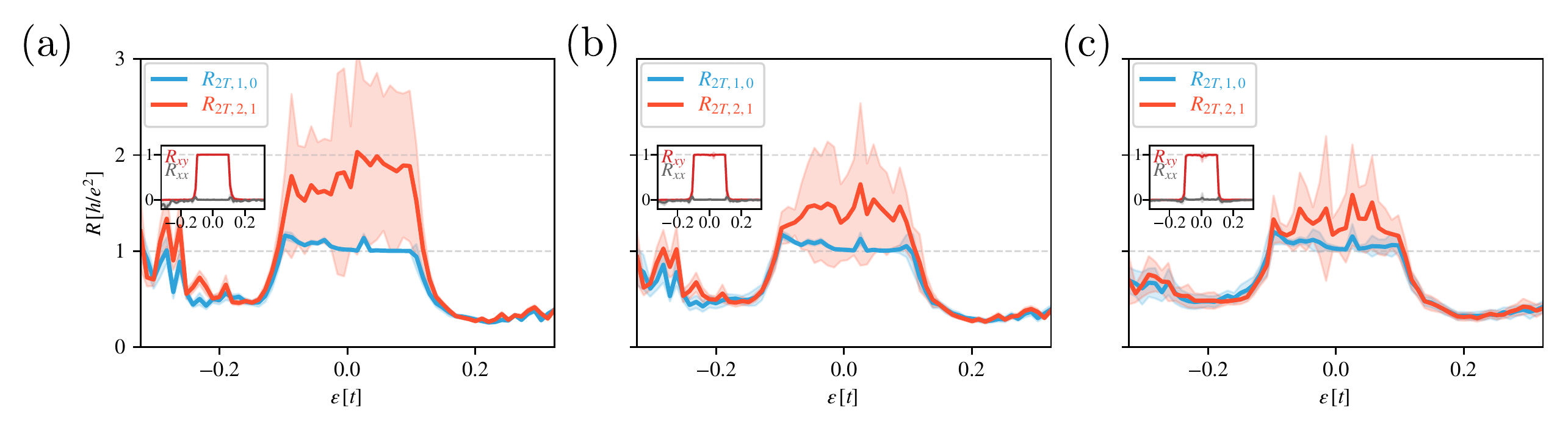}
    \caption{\label{struc_range}
    Resistance measurements simulations for slabs with structural disorder.
    The blue lines depict the matching case $R_{2T,1,0}$, while the red lines the mismatching case $R_{2T,2,1}$.
    The inset shows the typical non-local resistance measurements $R_{xx}$ and $R_{xy}$.
    The shaded region around the lines depict the standard deviation taken from $16$ disorder realizations. The panels (a), (b) and (c) correspond to $p=0.01$, $p=0.02$, and $p=0.05$, respectively.}
\end{figure}

To test the robustness of the HSP we have performed simulations with structural
disorder. We remove atomic sites randomly near the edges of the slab, in the
closest three unit-cells near the walls, with a probability $p$, as shown in
Fig.~\ref{struct_show}.
Note that we remove sites from a within $3$ lattice constants from the
edge. Since we remove each site with the same probability, for example
$\%5$, then the probability at any point on the side wall that at least one of
the sites is removed is much larger, around $\%14$.
We find that the \emph{spin-valve} effect is distinguishable with small
structural disorder (Fig.~\ref{struc_range}), and survives up to a structural disorder of $p=5\%$. While the two-terminal resistance in the matching case
remains quantized to the Hall resistance value, the mismatching case
shows a clear deviation from quantization that is dependent on the details
of the disorder realization. In Fig.~\ref{struc_range} we show the averaged
results for $16$ disorder realization, with shaded regions around the curves to
depict the standard deviation.

Other types of disorder, such as magnetic disorder, or a combination with
energetic disorder to mimic the effect of magnetic dopants,
are not simulated further in this Supplemental Material nor the main
Letter. However, we anticipate them to be
detrimental to the hinge spin polarization (HSP). Since we have demonstrated the
robustness of the HSP against energetic (Anderson) disorder and structural
disorder, we expect that there will also be a range of magnetic doping in which
the resistance signatures of the HSP are clearly discernible.

\end{document}